# Deep Learning Assisted Raman Spectroscopy for Rapid Identification of 2D Materials

**Yaping Qi**[1,2,#,*], **Dan Hu**[1,#], **Zhenping Wu**[3], **Ming Zheng**[4], **Guanghui Cheng**[2], **Yucheng Jiang**[5,*], **Yong P. Chen**[1,2,6,7,*]

[1]Department of Engineering Science, Faculty of Innovation Engineering, Macau University of Science and Technology, Av. Wai Long, Macau SAR, 999078, China

[2]Advanced Institute for Materials Research (WPI-AIMR), Tohoku University, Sendai 980-8577, Japan

[3]State Key Laboratory of Information Photonics and Optical Communications & School of Science, Beijing University of Posts and Telecommunications, Beijing 100876, China

[4]School of Materials Science and Physics, China University of Mining and Technology, Xuzhou 221116, China

[5]Jiangsu Key Laboratory of Micro and Nano Heat Fluid Flow Technology and Energy Application, School of Physical Science and Technology, Suzhou University of Science and Technology, Suzhou, Jiangsu 215009, China

[6]Department of Physics and Astronomy and Elmore Family School of Electrical and Computer Engineering and Birck Nanotechnology Center and Purdue Quantum Science and Engineering Institute, Purdue University, West Lafayette, Indiana 47907, United States

[7]Institute of Physics and Astronomy and Villum Center for Hybrid Quantum Materials and Devices, Aarhus University, Aarhus-C, 8000 Denmark

#These authors contributed equally: Yaping Qi, Dan Hu

*Correspondence: ypqi@must.edu.mo; jyc@usts.edu.cn; yongchen@purdue.edu.

**Abstract**

Two-dimensional (2D) materials have attracted extensive attention due to their unique characteristics and application potentials. Raman spectroscopy, as a rapid and non-destructive probe, exhibits distinct features and holds notable advantages in the structural characterization of 2D materials. However, traditional data analysis of Raman spectra relies on manual interpretation and feature extraction, which are both time-consuming and subjective. In this work, we employ deep learning techniques, including classificatory and generative deep learning, to assist the analysis of Raman spectra of typical 2D materials. For the limited and unevenly distributed Raman spectral data, we propose a data augmentation approach based on Denoising Diffusion Probabilistic Models (DDPM) to augment the training dataset and construct a four-layer Convolutional Neural Network (CNN) for 2D material classification. Experimental results illustrate the effectiveness of DDPM in addressing data limitations and significantly improved classification model performance. The proposed DDPM-CNN method shows high reliability, with 100%classification accuracy. Our work demonstrates the practicality of deep learning-assisted Raman spectroscopy for high-precision recognition and classification of 2D materials, offering a promising avenue for rapid and automated spectral analysis.

**Keywords**: Deep Learning; Raman Spectroscopy; Convolutional Neural Network; 2D Materials; Data Augmentation; Denoising Diffusion Probabilistic Models

## 1. Introduction

Since the discovery of graphene in 2004, an ever-expanding family of two-dimensional (2D) materials has been discovered and explored [1, 2]. Due to their unique physical and chemical properties, 2D materials have garnered significant attention in the scientific community and [3] exhibited tremendous potential in an extensive range of applications, such as photovoltaics, catalysis, sensors, and medicine [4-6]. To investigate the diverse properties of 2D materials, it is essential to characterize their basic structures and composition.

Raman spectroscopy is commonly employed as a measurement technique for analyzing 2D materials, it has been widely used in analytical sciences due to its sensitivity and non-invasive nature [7-9]. However, conventional Raman spectroscopy analysis involves laborious efforts, and human intervention for data interpretation [10-12]. To address these challenges, there has been a growing interest in integrating machine learning techniques with Raman spectroscopy. In the industrial community, efficient analysis through Raman spectroscopy is crucial to, particularly in areas such as quality control in product manufacturing, where rapid and accurate assessment of a large quantity of materials are necessary to ensure high-quality standards. Moreover, the complexity of material structures in industrial applications, involving multiple materials, necessitates the use of Raman spectroscopy combined with machine learning for rapid analysis of material interfaces and stacking arrangements.

Machine learning has attracted escalating interest among researchers for its ability to analyze complex spectral data [13-15]. For instance, algorithms such as random forest, kernel ridge regression, and multi-layer perceptron have been applied in the study of 2D materials, such as characterizing, differentiating their monolayer films in $MoS_2$, and identifying the angles of twisted bilayer graphene [16-19]. Despite these advancements enhancing the efficiency of traditional Raman analysis, it is important to note that the conventional machine learning methods largely depend on manual preprocessing and feature engineering of spectra to achieve optimal performance [20]. To address these limitations, researchers tend to apply deep learning to assist in Raman spectroscopy analysis [13, 20]. A prime example is the application of convolutional neural networks (CNN), which have been successfully employed in high-speed Raman imaging for the rapid identification of carbon nanotubes [21]. Furthermore, CNN have demonstrated their efficacy in accurately identifying hundreds of mineral categories and in distinguishing spectra of materials that are highly similar in subtly different environments [22, 23].

Nevertheless, deep learning requires extensive datasets to optimize the network parameters and mitigate overfitting risks [24]. Data augmentation methods such as generative adversarial networks (GAN), have shown promise in reducing overfitting and improving the accuracy of classification algorithms [25-27]. However, some research observed that GAN trade diversity for fidelity to produce high-quality samples but cannot cover the whole distribution of features in abundant sample scenarios [28-30]. In response to this constraint, our study proposes a novel approach, integrating a denoising diffusion probabilistic model (DDPM) with a 1D CNN-based classifier [30]. This hybrid model aims to identify Raman spectroscopy of 2D materials efficiently and accurately, including distinct types of 2D materials and their stacked combinations.

In this research, we explore the fusion of classification-focused deep learning and generative deep learning methodologies for the identification of various 2D materials through Raman spectroscopy. In response to the challenge of limited and non-uniformly distributed experimental Raman data of 2D materials, we implement advanced data augmentation strategies to substantially expand the number of training samples. The expansion is crucial for enhancing the performance of classification algorithms. Considering the characteristic diversity of one material in Raman spectral features derived from varied substrates, we have constructed a DDPM based on ResNet for data augmentation. Subsequently, we constructed a four-layer CNN for the automatic classification of spectra. This approach holds the potential to streamline the experimental process, reduce human intervention, and facilitate automated analysis of Raman spectra of 2D materials.

## 2. Materials and overall framework
### 2.1 Raman spectral data
This article defines the task of identifying various categories of 2D materials as a multi-class classification problem. In this study, we utilize experimental Raman spectra of 2D materials provided by Prof. Jiang from the Jiangsu Key Laboratory of Micro and Nano Heat Fluid Flow Technology and Energy Application as the dataset. This dataset comprises a total of 594 Raman spectra for seven distinct 2D materials and three stacked combinations: Black phosphorus (BP), Graphene, Molybdenum disulfide ($MoS_2$), Rhenium disulfide ($ReS_2$), Tellurium (Te), Tungsten diselenide ($WSe_2$), Tungsten ditelluride ($WTe_2$), BP–$WSe_2$ stack ($S_1$), Te-$ReS_2$-$WSe_2$-Graphene stack ($S_2$), and Te-$WSe_2$-$WTe_2$ stack ($S_3$). It is noteworthy that the spectral feature peaks of these materials might exhibit a diverse range of variations, as the spectra of each material are obtained under more than one substrate. The composition of the Raman spectroscopy dataset for 2D materials is depicted in Table 1, providing information about the quantity of spectra.

Table 1. Statistics of Raman spectral dataset of 2D materials studied in this work.

| Materials | Quantity of spectra |
|---|---|
| BP | 35 |
| Graphene | 209 |
| $MoS_2$ | 8 |
| $ReS_2$ | 15 |
| Te | 270 |
| $WSe_2$ | 6 |
| $WTe_2$ | 28 |
| $S_1$ | 8 |
| $S_2$ | 7 |
| $S_3$ | 8 |
| Total: | 594 |

**2.2 Overall framework**

In practical applications, the weak signals of trace substances are often difficult to separate from the substrate background using Raman analysis techniques, which makes it challenging to observe the signal of the target substance and limits the amount of spectral data obtained from experiments [31]. To address such issue, our framework introduces a novel data augmentation approach for Raman spectroscopy-based 2D material classification, as illustrated in Figure 1. It primarily consists of the following two components:

1. Data Augmentation Module: To augment the training dataset, we employ DDPM to generate synthetic samples for each category of materials. This process generates diverse samples, resulting in a substantial number of independently and identically distributed samples based on the original Raman spectral dataset. The objective is to assist the classification model in accurately and efficiently identifying various types of 2D materials.

2. Data Classification Module: Combining the original samples with those generated by DDPM, the data classifier learns to determine the category of each sample. In our study, we construct a four-layer CNNto classify each sample into their respective categories and compared it with other commonly used methods such as Artificial Neural Networks (ANN), Random Forest (RF), Support Vector Machine (SVM), K-Nearest Neighbors (KNN), and Logistic Regression (LR).

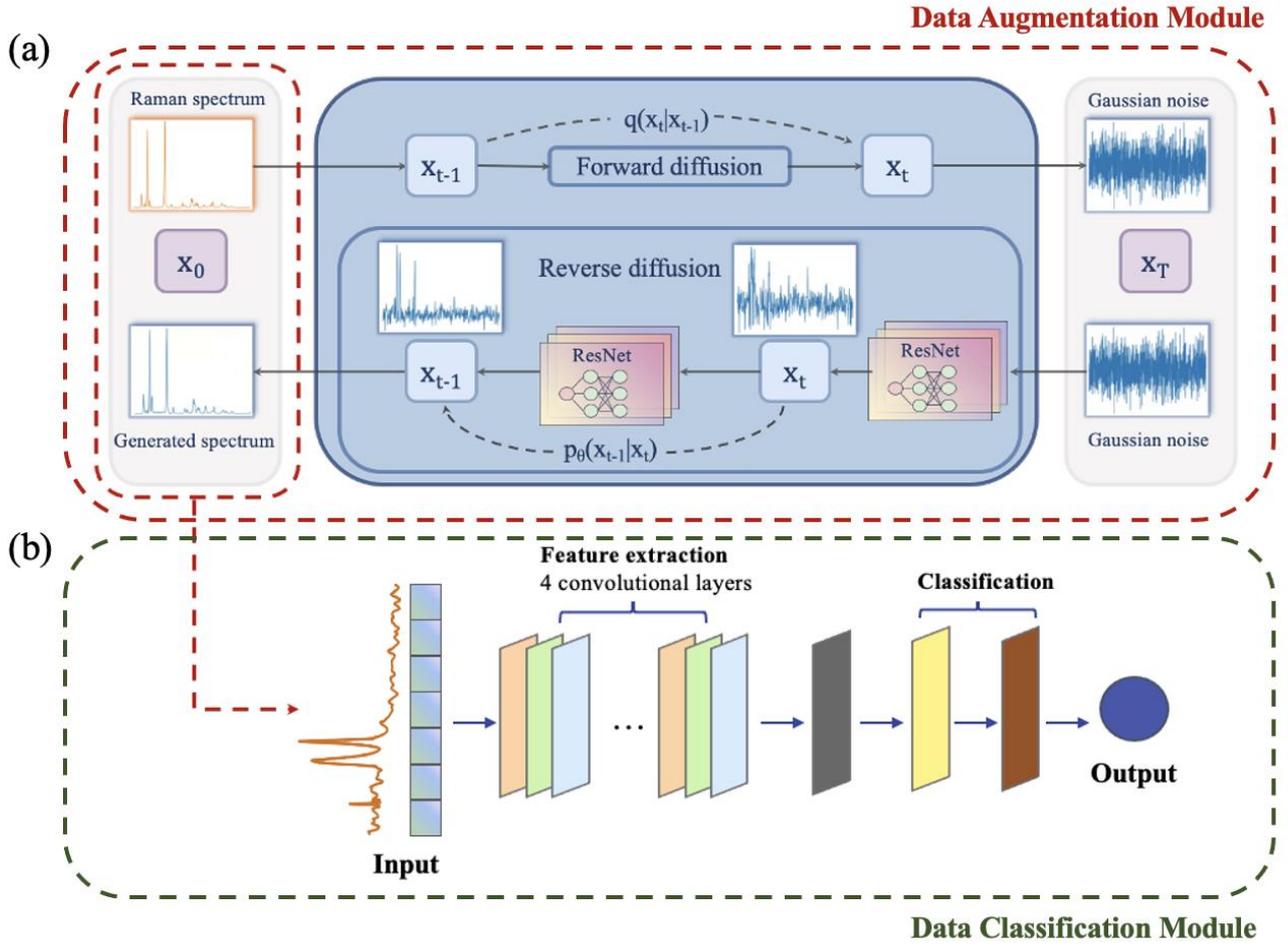

Figure 1. Illustration of the DDPM-based data augmentation for Raman Spectroscopy of 2D materials classification framework. (a). Data augmentation module based on DDPM. (b). Spectral classification module based on 1D CNN.

## 3. Methodology
### 3.1 Data augmentation module

Firstly, this study employs DDPMs for data augmentation. Earlier studies have used, DDPMs to synthesize high-quality data [32-35]. Diffusion probabilistic models were first introduced by Sohl-Dickstein et al. [36]. They defined a Markov chain of diffusion steps to construct desired data samples from noise by adding random noise to data and then learning to reverse the diffusion process.

Subsequently, Ho et al. (2020) proposed DDPM, a simplified diffusion model driven by the connection between denoising diffusion models and denoising fractional matching [37].

DDPM is composed of two processes: forward diffusion (right to left) and reverse diffusion (left to right), as shown in Figure 2.

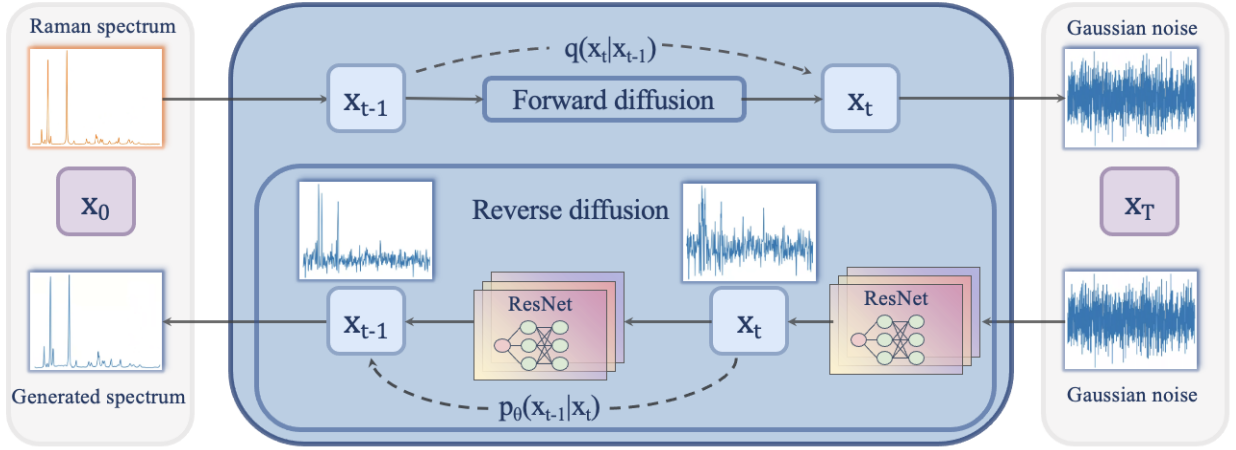

Figure 2. Illustration of DDPM-based data augmentation for Raman spectroscopy of 2D materials.

Forward diffusion is a process of adding noise to the input data, represented by $q$, it is fixed to a Markov chain from data $x_0$ to the latent variables $x_1, ..., x_T$:

$$q(x_1, ..., x_T | x_0) = \prod_{t=1}^{T} q(x_t | x_{t-1}) \quad (1)$$

The sampling noise latent based on the input $x_0$ at an arbitrary step can be expressed by defining $\alpha_t = 1 - \beta_t$ and $\overline{\alpha_t} = \prod_{t=0}^{T} \alpha_t$, where $\beta_1, ..., \beta_T$ are the noise schedule consisting of a set of linearly increasing constants:

$$x_t = \sqrt{\alpha_t} x_{t-1} + \sqrt{1 - \alpha_t} z_1 \quad (2)$$

$$x_t = \sqrt{\overline{\alpha_t}} x_0 + \sqrt{1 - \overline{\alpha_t}} z_t \qquad z_t \sim \mathcal{N}(0, I) \quad (3)$$

where $1 - \overline{\alpha_t}$ demonstrates the variance of noise for an arbitrary time step. Given sufficiently large time step T, the latent $x_T$ tends to the standard normal distribution $x_T \sim \mathcal{N}(0, I)$.

The reverse diffusion process is also defined as a Markov chain from the Gaussian noise input $x_T$ to $x_{T-1}, ..., x_0$. According to $q(x_T)$, we can sample the reverse steps $q(x_{t-1} | x_t)$. Here, we use $p_\theta$ to indicate the reverse process:

$$p_\theta(x_0, ..., x_{T-1} | x_T) = \prod_{t=1}^{T} p_\theta(x_{t-1} | x_t) \quad (4)$$

Using Bayes theorem, the diffusion process can be represented by the known quantities from the forward process, and it can be proved that $p_\theta(x_{t-1}|x_t, x_0)$ is also a Gaussian distribution:

$$p_\theta(x_{t-1}|x_t, x_0) = q(x_t|x_{t-1}, x_0) \frac{q(x_{t-1}|x_0)}{q(x_t|x_0)} \tag{5}$$

$$q(x_{t-1}|x_t, x_0) = \mathcal{N}(x_{t-1}; \tilde{\mu}_t(x_t, x_0), \tilde{\beta}_t I), \qquad \tilde{\beta}_t = \frac{1 - \overline{\alpha}_{t-1}}{1 - \overline{\alpha}_t} \beta_t \tag{6}$$

$$\tilde{\mu}_t(x_t, x_0) = \frac{\sqrt{\overline{\alpha}_{t-1}} \beta_t}{1 - \overline{\alpha}_t} x_0 + \frac{\sqrt{\alpha_t}(1 - \overline{\alpha}_{t-1})}{1 - \overline{\alpha}_t} x_t \tag{7}$$

The relationship between $x_0$ and $x_t$ is already obtained in the forward process:

$$\tilde{\mu}_t(x_t) = \frac{1}{\sqrt{\overline{\alpha}_t}} \left( x_t - \frac{1 - \alpha_t}{\sqrt{1 - \overline{\alpha}_t}} \overline{z}_t \right) \tag{8}$$

Since the noise $\overline{z}_t$ at time step t depends on the entire forward training process, it is hard to estimate. Therefore, we constructed residual networks (ResNet) based on the DiffWave model presented by Kong et al. (2020) to approximate the distribution of $\overline{z}_t$ in the reverse process. The structure of the ResNet is illustrated in Figure 3, it consists of eight residual layers and utilizes skip connections to connect the entire network.

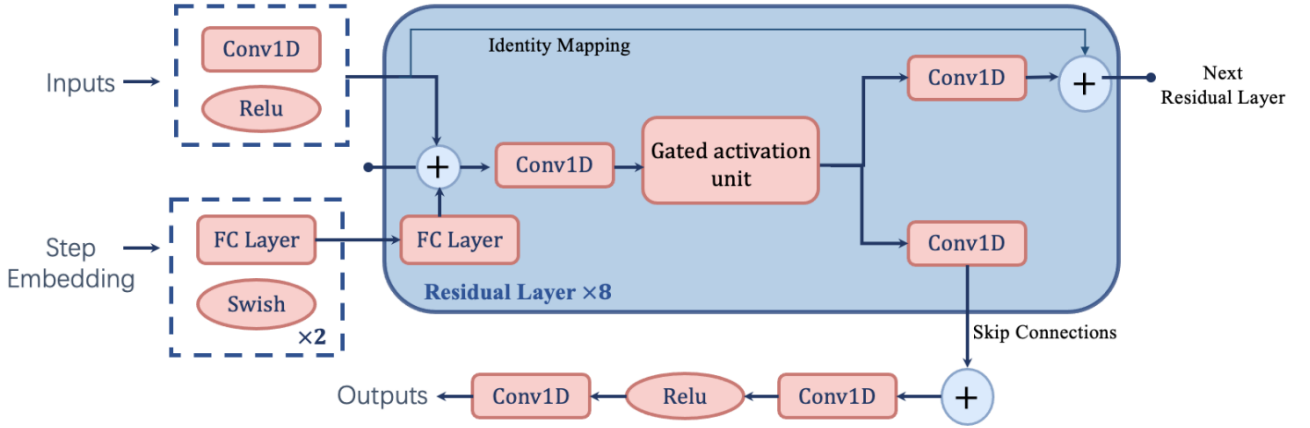

Figure 3. An illustration of the ResNets architecture. In this schematic, "FC" denotes the fully connected layer. "Relu" represents the rectified linear unit (Relu) activation function. The "gated activation unit" consists of the hyperbolic tangent (tanh) activation function and the sigmoid activation function, it can be denoted as $tanh(W_{f,k} * x) \odot \sigma(W_{g,k} * x)$, where W represents a convolutional filter, *f*, and *g* represent the filter and gate, respectively, and k represents the layer index.

The input of the model consists of two parts: input diffusion noise and step embedding, where the model needs to generate different diffusion results for different values of step t. Step embedding is a positional embedding introduced by Vaswani et al. [38], and in this study, we utilize it for the time step. The diffusion noise is fed into a 1D convolutional layer, while step t is input into a two-layer fully connected layer, where the parameters of these two parts are shared. Subsequently, step t is mapped to an embedding vector through the third fully connected layer, and together with the diffusion noise, it is added to the input of each residual layer in the model. Each residual layer utilizes one convolutional layer for feature extraction. The obtained features are then activated by gated activation units and passed through a pointwise convolutional layer. The output of the pointwise convolutional layer is divided into two parts along the channel dimension: one part is the input of the next residual layer, while the other is directly outputted through a skip connection, where the output module consists of two convolutional layers.

**3.2 Data classification module**

From the data augmentation module, we can obtain a set of new samples for each class of the original spectral data, which will be utilized to enhance the performance of the classifier. In our classification module, we employ 1D CNN as the core component. The neural network architecture constructed for classification is illustrated in Figure 4. The convolutional layer is crucial in CNNs for feature extraction. It convolves input data with trainable filters, directly influencing model performance. More convolutional layers allow the learning of additional features but increase training time. Each convolutional layer consists of trainable filters (kernels) that slide over input data, performing convolution operations. The process can be denoted as:

$$c_j^l = f\left(\sum_{i \in E_j} c_i^{l-1} * k_{ij}^l + b_j^l\right)$$

where * represents the convolution operation, $l$ denotes the current convolutional layer, $c_j^l$ is the output of $j_{th}$ feature map, $k_{ij}^l$ is the convolutional kernel, $E_j$ represents the input feature maps, f is the activation function, and b is the bias. The convolutional kernel hyperparameters are randomly initialized and optimized iteratively for optimal performance.

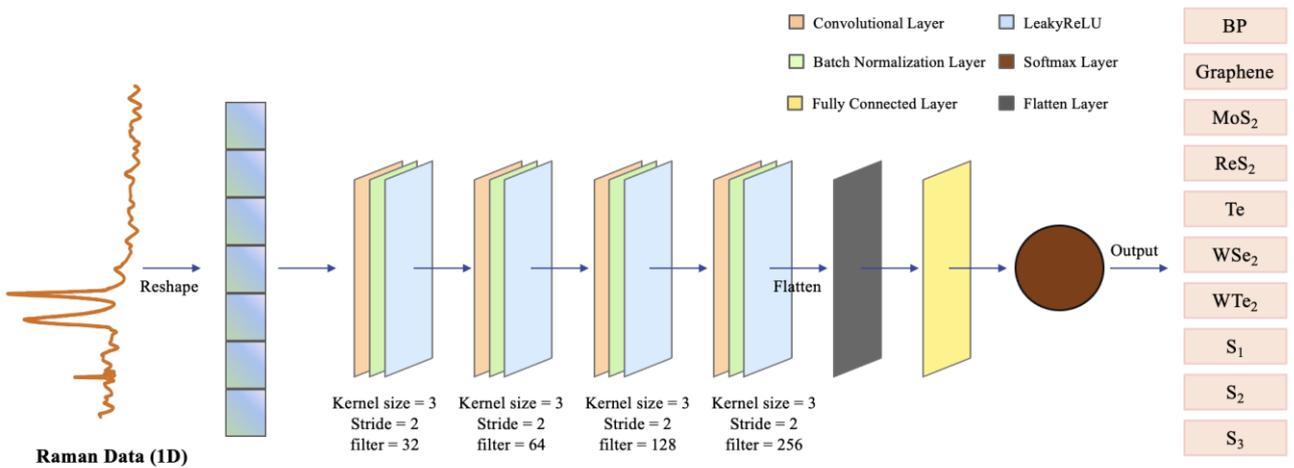

Figure 4. The architecture of the four-layer CNN for Raman spectroscopy classification.

Our CNN model uses four convolutional layers to extract data features and uses LeakyReLu as the activation function. Subsequently, a flattened layer is applied to transform the multi-dimensional input data into a set of 1D vectors, which is then fed into a fully connected layer. The fully connected layer receives the output from the convolutional and pooling layers and maps the learned features to a predefined vector space for feature classification. The expression for the fully connected layer is as follows:

$$h_{w,b}(x) = f(w^T x + b)$$

Where $h$ represents the output of the current neuron, x denotes the 1D feature vector input, and w corresponds to the weight vector connected to the neuron. Finally, there is a Softmax function with an output dimension equal to the number of classes. The Softmax function takes a set of 1D vectors as input and normalizes them into a probability distribution.

The classifier uses sparse categorical cross-entropy to calculate loss, which is expressed as follows:

$$loss = -\frac{1}{m}\sum_{i=1}^{m}\sum_{j=1}^{k} y_{ij} \log \hat{y}_{ij} \tag{9}$$

where m denotes the number of samples and denotes the number of categories, $y_{ij}$ means the real label of Raman data (if sample i belongs to class j then $y_{ij}$ is 1, else 0) and $\hat{y}_{ij}$ means the probability of model predicts the sample i belongs to class j.

## 4. Experiments and results
### 4.1 Data preprocessing
To ensure consistent dimensionality for the input of the model, we employ a simple spline interpolation technique to convert each Raman spectrum into a vector of 571 intensity values, within the wavenumber range of 50-1750 cm$^{-1}$. This range is selected to maximize the information within each spectrum, and effectively encompass the characteristic peaks necessary for differentiating the Raman spectra of 2D materials. For spectra that do not cover the entire range of wavenumbers, the missing intensity values were padded with zeros. Finally, the dataset is normalized to establish consistent scaling across all features, thus preparing for model training and analysis.

### 4.2 Implementation settings
The experiments in this study are conducted on the Windows 11 operating system using Python 3.9 programming language. The hardware used for the experiments includes a CPU of 12th Gen Intel(R) Core (TM) i7-12700 and a GPU of Nvidia GeForce RTX 3060Ti.

In the data augmentation module during the training of ResNet, each convolutional layer has a kernel size of 3, and the channel dimension within the residual blocks is set to 128. The generation of diffusion noise follows a linear schedule spanning 50 steps, with the range of $\beta_t$ values set between 0.0001 and 0.02. For the proposed four-layer CNN in the classification module, the kernel size is set to 3, and a stride of 2 is applied. The number of filters is configured as 32, 64, 128, and 256 for the respective layers. All neural network models are optimized using the Adam optimizer with an initial learning rate of 0.0002. The model undergoes training for 100 epochs, utilizing a batch size of 32. The training and test datasets for all models are divided in a 4:1 ratio.

**4.3 Generated data**

To ensure a comprehensive augmented dataset, the experiment utilize the best-saved model to generate Raman spectra for each trained diffusion model. Figure 5 shows the Raman spectra of ten categories of materials included in the dataset, as well as the generated Raman spectra. We employ DDPM to augment 1000 spectral data for each type of 2D material, generating a total of 10,000 Raman spectrum samples for further analysis. It can be observed that DDPM can generate diverse synthetic spectra that closely resemble the features of the original spectra. Additionally, DDPM exhibits the ability to fill in new data within a specific range based on original data (extrapolate data within predefined limits using the original data as a reference). This capability enables the comprehensive capture of all characteristics and improves the diversity of the dataset.

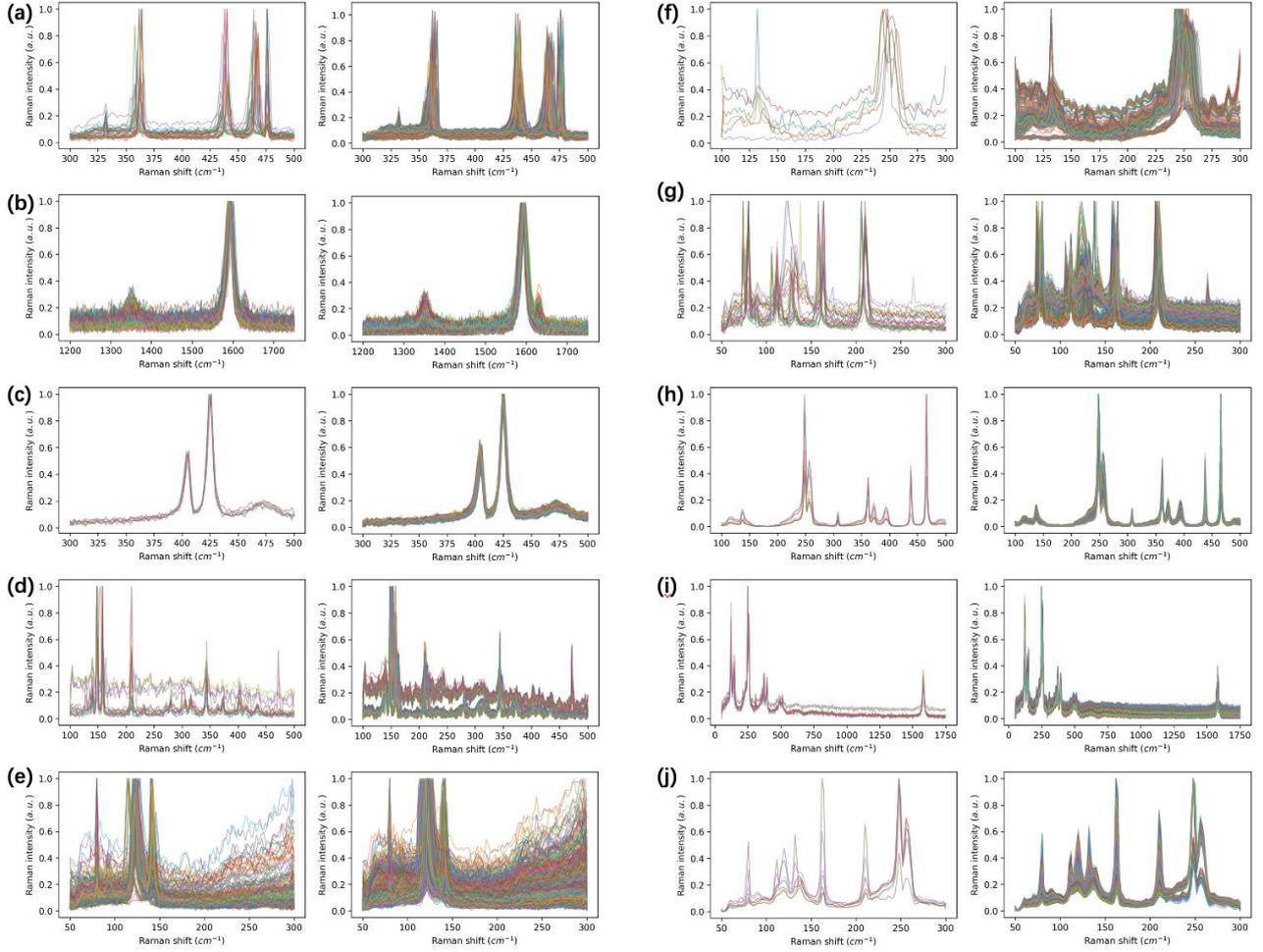

Figure 5. The 2D material Raman spectra before and after data augmentation using DDPM: (a) BP, (b) Graphene, (c) MoS$_2$, (d) ReS$_2$, (e) Te, (f) WSe$_2$, (g) WTe$_2$, (h) BP–WSe$_2$ stack, (i) Te-ReS$_2$-WSe$_2$-Graphene stack, and (j) Te-WSe$_2$-WTe$_2$ stack. The left side represents the original Raman spectra dataset, while the right represents the augmented Raman spectra dataset.

Furthermore, we employ the t-SNE dimensionality reduction technique (Figure 6) to visualize the original data and new samples generated by the DDPM. It can be observed that the features of Raman spectra generated by DDPM for different categories exhibit distinct boundaries in the low-dimensional space. Therefore, integrating the generated spectra into the dataset facilitates more diverse and comprehensive analysis, enabling a robust evaluation of deep learning-assisted methods.

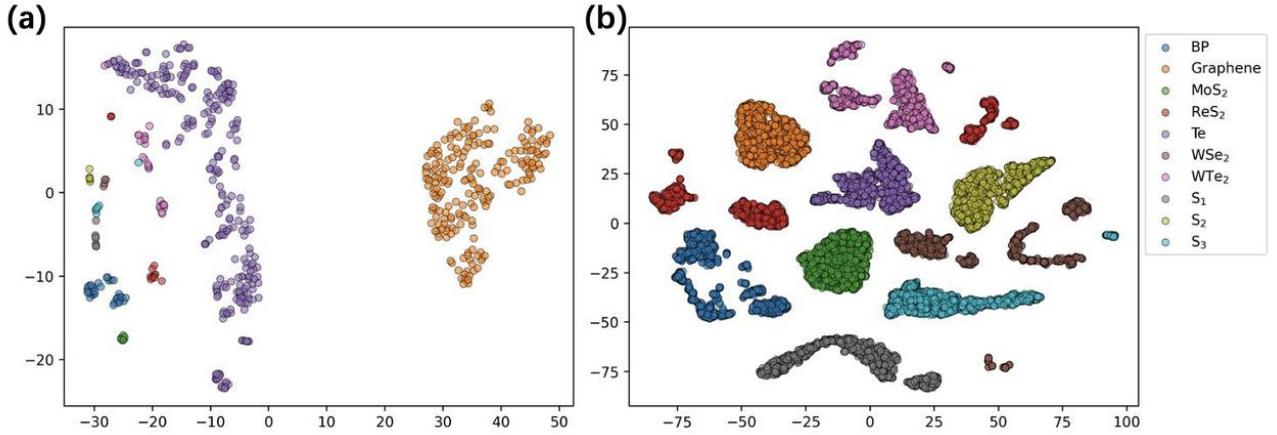

Figure 6. t-SNE plots for (a) the original and (b) the augmented dataset of different 2D materials. [$S_1$: BP–$WSe_2$ stack, $S_2$: Te-$ReS_2$-$WSe_2$-Graphene stack, $S_3$: Te-$WSe_2$-$WTe_2$ stack.]

**4.4 Results and analysis**

To assess the advantages of the proposed model, the experiments conduct the following baselines for comparison: RF, SVM, KNN, LR, and an ANN model with two hidden layers. For the multi-class classification task to identify 2D materials, we used the average accuracy, precision, and recall of ten-fold cross-validation as evaluation metrics.

Table 2 reports the performance comparisons between the proposed method against baselines. It is worth noting that CNN and DDPM-CNN exhibit superior classification performance compared to other models in the evaluation. Specifically, CNN achieves an exceptional accuracy rate of 98.8% without data augmentation, surpassing most other models. This indicates its ability to accurately classify data and exhibit good generalization. Figure 7 supplements this evaluation with an array of confusion matrices for different algorithms, where CNN displays higher accuracy in classifying most categories. Such results corroborate that deep learning methods can effectively extract sample features even when training data is limited. Conversely, conventional machine learning techniques such as RF and KNN tend to underperform when relying on raw Raman data as input features, which may limit the exploitation of inter-feature correlations, thereby affecting classifier performance. Moreover, with a precision of 94.5% and a recall of 93.7%, it indicates that CNN still faces challenges in accurately identifying positives and capturing all True Positive instances.

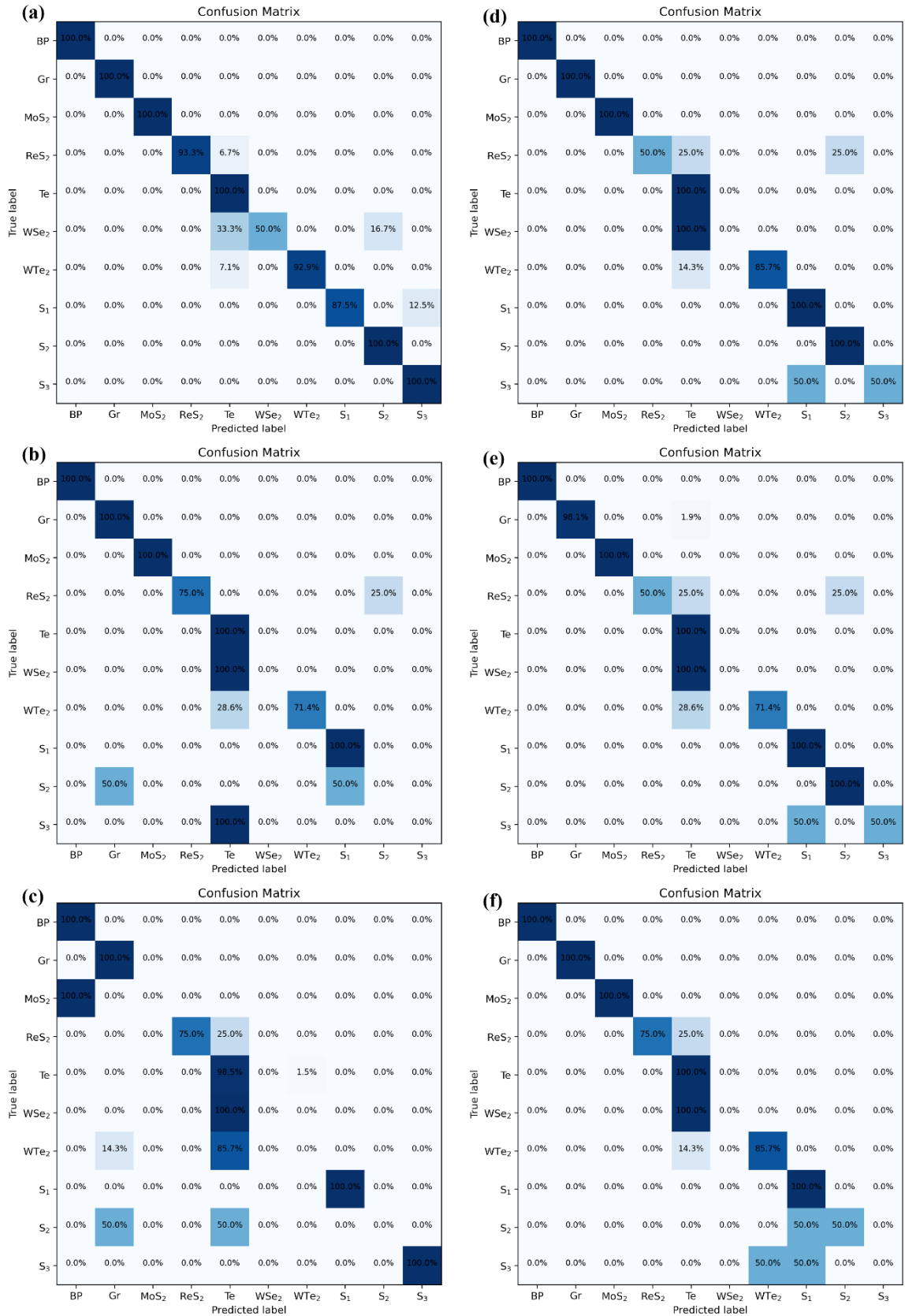

Figure 7. Confusion matrices depicting the average accuracy of ten-fold cross-validation in the classification of each category by algorithms: (a) CNN, (b) ANN, (c) RF, (d) SVM, (e) KNN, and (f) LR. The diagonal elements represent the percentage of true positives, which is a key indicator of the algorithm's ability to correctly identify each category. The off-diagonal elements represent misclassification rates.

In comparison, models such as ANN, RF, SVM, KNN, and LR, demonstrate varying levels of performance in accuracy, precision, and recall. Although these models may slightly be inferior to CNN, their performances nonetheless indicate their capabilities for 2D material recognition. It is worth emphasizing that the incorporation of the DDPM as a data augmentation method significantly enhances the performance across all evaluated models. Notably, the average accuracy in ten-fold cross-validation of DDPM-ANN and DDPM-RF models ascended from 94.6% and 90.6% to 100%. This highlights the effectiveness of DDPM in refining algorithm performance in 2D material recognition.

However, the advantages of DDPM-CNN are not prominent due to significant differences among the data categories used in this study. All DDPM-based conventional machine learning can achieve remarkable results. Typically, deep neural networks require more data to reach their optimal performance, so further validation of its performance can be conducted on more complex datasets with smaller inter-category differences (difficult to distinguish).

Table 2: The average performance of ten-fold cross-validation comparisons between the proposed method vs. baselines.

| Method | Accuracy | Precision | Recall |
|---|---|---|---|
| **CNN** | 0.988 | 0.945 | 0.937 |
| **DDPM-CNN** | 1.000 | 1.000 | 1.000 |
| **ANN** | 0.946 | 0.658 | 0.646 |
| **DDPM-ANN** | 1.000 | 1.000 | 1.000 |
| **RF** | 0.906 | 0.566 | 0.574 |
| **DDPM-RF** | 1.000 | 1.000 | 1.000 |
| **SVM** | 0.966 | 0.829 | 0.786 |
| **DDPM-SVM** | 1.000 | 1.000 | 1.000 |
| **KNN** | 0.953 | 0.826 | 0.770 |
| **DDPM-KNN** | 0.988 | 0.989 | 0.988 |
| **LR** | 0.960 | 0.731 | 0.711 |
| **DDPM-LR** | 1.000 | 1.000 | 1.000 |

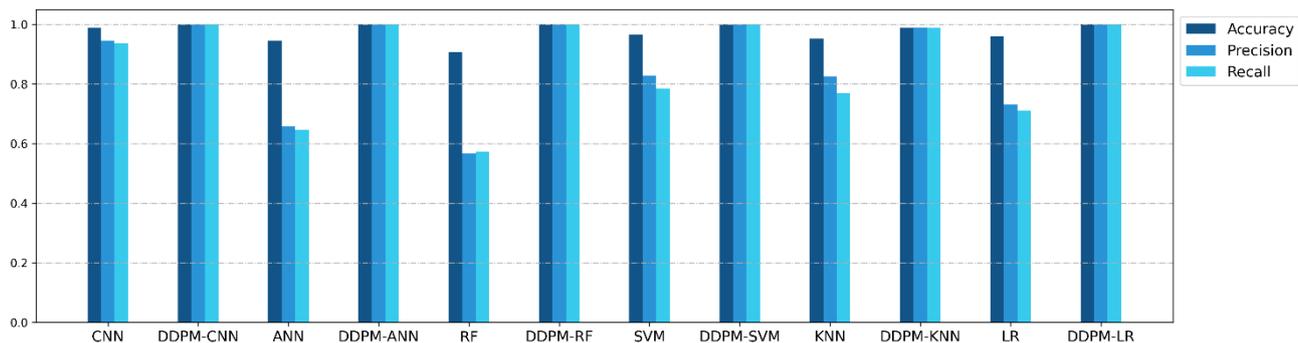

Figure 8: Bar chart of the average performance of ten-fold cross-validation between the proposed method vs. baselines.

The data augmentation module based on DDPM proposed in this study significantly enhances sample density and diversity, enabling the classifier to establish decision boundaries more effectively. As a result, it outperforms baseline models. Higher sample density, in comparison to sparse data, often allows classifiers to learn more precise boundaries. Overall, the utilization of DDPM-based data augmentation has the potential to be a valuable technique in materials science. Its ability to generate realistic spectra and improve the recognition capabilities of classification models. The findings underscore the effectiveness of leveraging data augmentation methods for more accurate and robust 2D material recognition, ultimately contributing to the progress and exploration of novel materials in the scientific community.

## 5. Conclusion

This study explores the application of deep learning techniques to assist in identifying different 2D materials based on Raman spectroscopy. In response to the challenge of limited data availability, we employ data augmentation techniques to substantially augment the training samples to improve the effectiveness of the classification. We have constructed a DDPM-based augmentation model with ResNet, which effectively addresses data distribution, promotes diversity, and boosts the performance of all classification models, including CNN, ANN, RF, SVM, KNN, and LR. The four-layer CNN model that we constructed demonstrates exceptional performance in this study, achieving classification accuracies of 98.8% without data augmentation and a score of 100% upon integrating DDPM-based data augmentation. These outcomes highlight the practicality of the proposed data augmentation approach, enabling high-precision identification of 2D materials even in small-scale data tasks. Furthermore, this study is the inaugural application of the DDPM in spectral generation, presenting a novel tool for data augmentation in Raman spectroscopy and other spectral analysis. It can simplify the experimental process, reduce human intervention, and facilitate automated analysis of spectroscopy, thus paving a new avenue for further research in this domain.

**Data availability**

The data are available from the corresponding authors upon reasonable request.

**Code availability**

https://github.com/Naduhi/Raman_DL

**Acknowledgments**

We acknowledge partial support of this work by Macau Science and Technology Development Fund (FDCT Grants 0031/2021/ITP), and Purdue University Discovery Park Big Idea Challenge program. We also thank the support by Tohoku University TUMUG Startup Research Fund, and AIMR Overseas dispatch program for young researchers FY2023.

**Conflict of interest**

The authors declare no conflict of interest.

**CRediT author statement:**

**Yaping Qi:** Conceptualization, Investigation, Methodology, Project Administration, Resources, Funding acquisition, Software, Validation, Supervision, Writing-Original draft, Writing-reviewing and Editing. **Dan Hu:** Investigation, Methodology, Software, Validation, Writing-Original draft, Writing-reviewing and Editing. **Zhenping Wu:** Resources, Writing-Original draft, Writing-reviewing and Editing. **Ming Zheng:** Resources, Writing-Original draft, Writing-reviewing and Editing. **Guanghui Cheng:** Resources, Writing-Original draft, Writing-reviewing and Editing. **Yucheng Jiang:** Conceptualization, Project Administration, Resources, Data Curation, Writing-Original draft, Writing-reviewing and Editing. **Yong P. Chen:** Conceptualization, Project Administration, Resources, Funding acquisition, Supervision, Writing-reviewing and Editing.